\documentclass[a4paper,11pt]{article}
\usepackage{pos}

\newcommand{\be}{\begin{equation}}
\newcommand{\ee}{\end{equation}}
\newcommand{\ba}{\begin{eqnarray}}
\newcommand{\ea}{\end{eqnarray}}
\newcommand{\ban}{\begin{eqnarray*}} 
\newcommand{\ean}{\end{eqnarray*}}
\newcommand \nn {\nonumber}

\def\P{{\mathbf P}}
\def\b{{\mathbf b}}

\def\k{{\mathbf k}}

\def\r{{\mathbf r}}
\def\z{{\mathbf z}}
\def\w{{\mathbf w}}
\def\v{{\mathbf v}}

\title{Back-to-back dijet production in DIS with finite-energy corrections and twist-3 gluon TMDs}

\author[a]{Tolga Altinoluk}
\author[a]{Guillaume Beuf}
\author[a]{Alina Czajka}
\author[b]{Cyrille Marquet}

\affiliation[a]{Theoretical Physics Division, National Centre for Nuclear Research,\\ Pasteura 7, Warsaw 02-093, Poland}

\affiliation[b]{CPHT, CNRS, \'Ecole polytechnique, Institut Polytechnique de Paris,\\ 91120 Palaiseau, France}

\emailAdd{tolga.altinoluk@ncbj.gov.pl}
\emailAdd{guillaume.beuf@ncbj.gov.pl}
\emailAdd{alina.czajka@ncbj.gov.pl}
\emailAdd{cyrille.marquet@polytechnique.edu}

\abstract{This work presents the summary of calculation of the cross section of the dijet production in deep inelastic scattering at small x at next-to-eikonal accuracy. The cross section is calculated in the back-to-back limit of the produced jets using results obtained in our previous works. The cross section is expressed via the transverse-momentum-dependent (TMD) parton distributions. Specifically, we show how the next-to-eikonal corrections are related to the $x$ dependent phase of twist-2 gluon TMD and to twist-3 unpolarized gluon TMDs.}

\FullConference{Proceedings of the Corfu Summer Institute 2025 "School and Workshops on Elementary Particle Physics and Gravity" (CORFU2025)\\
27 April - 28 September, 2025\\
Corfu, Greece\\}

\begin{document}
\maketitle

\section{Introduction}

Deep inelastic scattering (DIS) on a dense gluon target is one of the processes used to study gluon saturation effects of the target. It provides a clean environment for such studies and therefore will be at the core of experimental focus in the Electron Ion Collider (EIC). In that process an incoming lepton emits a virtual photon, which then splits into a pair of quark and antiquark. The pair interacts perturbatively multiple times when traversing the target and the scattering is treated within the Color Glass Condensate (CGC) approach. For reviews of CGC see, for example, \cite{Gelis:2010nm,Albacete:2014fwa}. The observables produced in DIS processes are studied using the dipole factorization framework~\cite{Bjorken:1970ah,Nikolaev:1990ja}.

While the calculations of DIS related observables have been mostly focused on including NLO corrections, the newly-built EIC has stimulated the interest in effects occurring at relatively low energies where next-to-eikonal corrections become important. Studies on next-to-eikonal corrections involve relaxation of shockwave approximation, which is a typical approximation of a target at high energies. Therefore, by going beyond the shockwave approximation one relaxes the Lorentz contraction in the longitudinal direction, relaxes the Lorentz time dilation to include dynamics of the target and includes the subleading (to "+" component) transverse component of the gauge field. Over last years, there have been significant improvements in calculation of subeikonal corrections, see for example \cite{Altinoluk:2014oxa,Altinoluk:2015gia,Altinoluk:2015xuy,Agostini:2019avp,Agostini:2019hkj,Agostini:2022ctk,Agostini:2022oge,Altinoluk:2020oyd,Agostini:2023cvc,Altinoluk:2021lvu,Altinoluk:2022jkk,Agostini:2024xqs,Altinoluk:2023qfr,Chirilli:2018kkw,Chirilli:2021lif}.

In Ref.~\cite{Dominguez:2011wm} it was observed that in a specific kinematic limit, which is known as correlation limit, results on dihadron or dijet production obtained within CGC can be connected to the high energy limit of the leading twist gluon Transverse Momentum Dependent distribution functions (TMDs). These findings triggered multiple efforts to explore the connection between CGC and TMD factorization frameworks, see for example \cite{Altinoluk:2019fui,Altinoluk:2019wyu,Boussarie:2020vzf,Mukherjee:2026cte}.

In this work we present a summary of DIS dijet production calculation with next-to-eikonal corrections, which is done within CGC approach in the correlation limit, and then study its connection to TMD factorization framework beyond leading twist. We focus only on the case when the photon is longitudinally polarized. All details of the calculation, the case of the transverse photon and more comprehensive discussions are shown in our original paper \cite{Altinoluk:2024zom}.

\section{Definition and notation}

The cross section at next-to-eikonal accuracy for DIS dijet for longitudinal photon is defined as \cite{Altinoluk:2022jkk}
%
\begin{align}
 \frac{d\sigma_{\gamma^{*}_L\rightarrow q_1\bar q_2}}{d {\rm P.S.}} =  \frac{d\sigma_{\gamma^{*}_L\rightarrow q_1\bar q_2}}{d {\rm P.S.}}\Bigg|_{\rm Gen. \, Eik}+  \frac{d\sigma_{\gamma^{*}_L\rightarrow q_1\bar q_2}}{d {\rm P.S.}}\Bigg|_{\rm NEik \, corr.} + O({\rm NNEik})
 \label{def_L_cross_sec}
\end{align}
%
with the Lorentz invariant phase space measure for the produced dijets given by
\begin{align}
d{\rm P.S.}=\frac{d^2\k_1}{(2\pi)^2}\frac{dk_1^+}{(2\pi)2k_1^+}\frac{d^2\k_2}{(2\pi)^2}\frac{dk_2^+}{(2\pi)2k_2^+}\, .
\label{phase_space_1_2}
\end{align}
The momenta of produced jets (quark and antiquark) are denoted by $k_1$ and $k_2$. The first contribution to the cross section in Eq.~(\ref{def_L_cross_sec}) is the generalized eikonal cross section given by
%
\begin{align}
\label{Cross_Section_GEik}
\frac{d\sigma_{\gamma^{*}_L\rightarrow q_1\bar q_2}}{d {\rm P.S.}}\Bigg|_{\rm Gen. \, Eik} \!\!\!
= 2q^+ \!\! \int \!\! d (\Delta b^-) e^{i\Delta b^-(k_1^++k_2^+ - q^+)} 
\!\!\!
\sum_{\rm hel. \,,\,   col. }
\!\!
\Big\langle  \Big(\mathbf{M}_{q_1 \bar q_2 \leftarrow \gamma^*_L}^{\rm Gen.\,  Eik} \Big(- \frac{\Delta b^-}{2}\Big)\Big)^\dag \mathbf{M}_{q_1 \bar q_2 \leftarrow \gamma^*_L}^{\rm Gen.\,  Eik} \Big(\frac{\Delta b^-}{2} \Big) \Big\rangle
\, ,
\end{align}
%
where $q^+$ is the momentum of the photon and $\mathbf{M}_{q_1 \bar q_2 \leftarrow \gamma^*_L}^{\rm Gen.\,  Eik}$ is the generalized eikonal scattering amplitude. The sum is taken over colors and helicities of the produced jets, and $\langle \dots \rangle$ denotes target averaging. The generalized eikonal term goes beyond the strict eikonal result due to weak dependence on the light-cone coordinate $z^-$ of the background gluon field, which represents relaxing infinite Lorentz time dilation. The strict eikonal contribution can be obtained by neglecting $b^-$ dependence in the formula~(\ref{Cross_Section_GEik}).

The second term in Eq.~\eqref{def_L_cross_sec} contains the explicit next-to-eikonal corrections to the cross section and is given by 
\begin{align}
 \frac{d\sigma_{\gamma^{*}_L\rightarrow q_1\bar q_2}}{d {\rm P.S.}}\Bigg|_{\rm NEik \, corr.}
&
=  (2q^+)\, 2\pi \delta(k_1^+\!+\!k_2^+\!-\!q^+) \sum_{\rm hel. \,,\,  col. }
2 {\rm Re}  \Big\langle  \Big({\cal M}_{q_1 \bar q_2 \leftarrow \gamma^*_L}^{\rm strict~Eik}\Big)^\dag  {\cal M}_{q_1 \bar q_2 \leftarrow \gamma^*_L}^{\rm NEik \, corr.} \Big\rangle
\, ,
\label{Cross_Section_long_NEik}
\end{align}
where the next-to-eikonal correction to the amplitude is the sum of the following three different contributions:
\begin{align}
i {\cal M }_{q_1 \bar q_2 \leftarrow \gamma^*_L}^{\rm NEik \, corr.}= 
i{\cal M }_{q_1 \bar q_2 \leftarrow \gamma^*_L}^{\textrm{dec. on }q}
 +
i{\cal M }_{q_1 \bar q_2 \leftarrow \gamma^*_L}^{\textrm{dec. on }\bar{q}}
+
i{\cal M }_{q_1 \bar q_2 \leftarrow \gamma^*_L}^{\textrm{dyn. target}}
\, ,
\label{NEik_corr_Ampl_L_m}
\end{align}   
where the first and the second term stand for amplitudes calculated with decorated quark and antiquark, respectively, and the last term corresponds to the contribution with $z^-$ dependence included. 

Our goal is to study the DIS dijet production in the correlation limit, where it is possible to track the interplay of CGC and TMD factorization approaches. In this limit, it is useful to change variables for the generic quark and antiquark momenta $(\k_1,\k_2) \to (\P,\k)$ so that 
\ba
\k_1 &=& \P + z_1 \k, \\
\k_2 &=& -\P + z_2 \k
\, ,
\ea
where we introduced
\ba
z_{1,2} &\equiv& \frac{k^+_{1,2}}{k_1^+\!+\!k_2^+} .
\label{def_z1_z2}
\ea
Similarly, the transverse positions are also changed $(\v,\w) \to (\r,\b)$ as follows
\ba
\v &=& \b +z_2 \r  
\label{def_v_from_r_b}\\
\w&=& \b-z_1 \r
\label{def_w_from_r_b}
\, .
\ea
In the new variables $\k$ is the total momentum of the jets, $\P$ is the relative transverse momentum between the jets, $\r$ is the dipole size and $\b$ is the dipole impact parameter. The back-to-back jets kinematics shall be identified with taking the limits: $|\k| \ll |\P|$ and $|\r| \ll |\b|$. This regime is particularly convenient as it allows to expand color operators as powers of $\r$ and systematically study the corrections of the type $(|\k|/|\P|)^n$.

In the correlation limit the formula~(\ref{NEik_corr_Ampl_L_m}) can be reorganized to the form
\begin{align}
i {\cal M }_{q_1 \bar q_2 \leftarrow \gamma^*_L}^{\rm NEik \, corr.}= 
i{\cal M }_{q_1 \bar q_2 \leftarrow \gamma^*_L}^{(1)+(2)}
 +
i{\cal M }_{q_1 \bar q_2 \leftarrow \gamma^*_L}^{(3)}
+
i{\cal M }_{q_1 \bar q_2 \leftarrow \gamma^*_L}^{\textrm{dyn. target}}
\, ,
\label{NEik_corr_Ampl_L_b2b}
\end{align}   
where the first term comes from contributions involving derivatives of the Wilson lines and it contains only $\mathcal{F}_{i}^{\,-}$ components of the background field, the second term involves the component $\mathcal{F}^{ij}_a$, and the third term is given by $\mathcal{F}^{+-}$.

In these proceedings we do not present explicit forms of the contributions to the scattering amplitude in general kinematics and we do not discuss all possible structures emerging at subeikonal order. All these issues are comprehensively discussed in our previous papers~\cite{Altinoluk:2024zom,Altinoluk:2022jkk,Altinoluk:2020oyd}.

\section{The amplitude of the back-to-back dijet production in DIS}

The scattering amplitude of the longitudinal photon splitting into quark-antiquark pair in the correlation limit in the strict eikonal approximation is given by
\begin{align}
& i{\cal M }_{q_1 \bar q_2 \leftarrow \gamma^*_L}^{\rm   Eik} 
=\, 
 - \frac{Q\,  e e_f}{q^+}\, \bar u(1) \gamma^+ v(2)\, 
z_1 z_2\, 
\int_{\b}\,   e^{-i\b\cdot\k}\,
\nn \\
&
\times
\Bigg\{
\bigg[
-\frac{2\P^j}{[\P^2\!+\!\bar Q^2]^2}
+(z_2\!-\!z_1)\bigg(\frac{\k^j}{[\P^2\!+\!\bar Q^2]^2}
-\frac{4(\k\!\cdot\!\P)\P^j}{[\P^2\!+\!\bar Q^2]^3}
\bigg)
\bigg]\,
t^{a'} 
\!\!\!\int  dv^+ 
\mathcal{U}_A\left(+\infty,v^+;\b\right)_{a'a} g{\mathcal{F}_{j}^{\;-}}_a (v^+,\b)\; 
\nonumber\\
&
+\bigg[
-\frac{\delta^{ij}}{[\P^2\!+\!\bar Q^2]^2}
+\frac{4\P^i\P^j}{[\P^2\!+\!\bar Q^2]^3}
\bigg]\, t^{a'} t^{b'}
\int  dv^+ \int  dw^+ 
\, \mathcal{U}_A\left(+\infty,v^+;\b\right)_{a'a} g{\mathcal{F}_{i}^{\;-}}_a (v^+,\b)\; 
\nonumber\\
&
\times
\mathcal{U}_A\left(+\infty,w^+;\b\right)_{b'b} g{\mathcal{F}_{j}^{\;-}}_b (w^+,\b)
\Bigg\}
+O\left(\frac{Q|\k|}{|\P|^5}\right)
\label{Ampl-Eik_L_b2b} 
\, .
\end{align}
and this contribution was found from the generalized eikonal amplitude after neglecting $z^-$ dependence in the color structures. In the formula~(\ref{Ampl-Eik_L_b2b}), $\mathcal{U}_A$ stands for the Wilson line in the adjoint representation, $\mathcal{F}_i^{\,-}$ is the leading component of the strength tensor of the background gluon field and ${\bar Q}= \sqrt{m^2+Q^2k_1^+k_2^+/(q^+)^2}$ with $Q$ being the photon virtuality.

In Eq.~\eqref{Ampl-Eik_L_b2b}, the term proportional to $-2\P^j$ is the leading-power contribution, which is of order $Q/(|\k|\, |\P|^3)$ at amplitude level. The other explicitly written terms are next-to-leading-power corrections, of order $Q/|\P|^4$. 

The first next-to-eikonal correction that comes from decorated Wilson lines involving one or two covariant derivatives, $\mathcal{U}^{(1)}_{F;j}$ and $\mathcal{U}^{(2)}_F$, is found to be of the form
\begin{align}
& i{\cal M }_{q_1 \bar q_2 \leftarrow \gamma^*_L}^{(1)+(2)}
=
 -  \frac{Q\,e e_f}{q^+}\, \bar u(1)  \gamma^+ v(2) \int_{\b}\,   e^{-i\b\cdot\k}\,
 \nn \\
& 
\times
\Bigg\{
\frac{i}{2q^+}\bigg[
\frac{-2\P^j}{[\P^2\!+\!\bar Q^2]}
+(z_2\!-\!z_1)\bigg(\frac{\k^j}{[\P^2\!+\!\bar Q^2]}
-\frac{4(\k\!\cdot\!\P)\P^j}{[\P^2\!+\!\bar Q^2]^2}
\bigg)
\bigg]\,
t^{a'} \!\!\! \int  \!dv^+\, v^+\, 
 \mathcal{U}_A\left(+\infty,v^+;\b\right)_{a'a} g{\mathcal{F}_{j}^{\;-}}_a (v^+,\b)\; 
\nonumber\\
&
+\frac{i}{2q^+}\, t^{a'} t^{b'}\int  dv^+ \int  dw^+
\mathcal{U}_A\left(+\infty,v^+;\b\right)_{a'a} g{\mathcal{F}_{i}^{\;-}}_a (v^+,\b)\; 
\mathcal{U}_A\left(+\infty,w^+;\b\right)_{b'b} g{\mathcal{F}_{j}^{\;-}}_b (w^+,\b)
\nonumber\\
&
\times
\Bigg[
-\frac{\delta^{ij}}{[\P^2\!+\!\bar Q^2]}\, \min(v^+,w^+)
+\frac{4\P^i\P^j}{[\P^2\!+\!\bar Q^2]^2}\, ( z_2 v^+\!+\!z_1 w^+)
\Bigg]
\Bigg\}+O\left(\frac{Q |\k|}{W^2\, |\P|^3}\right)
\label{ampl-1_2_L_b2b_bis} 
\, .
\end{align}
The next subeikonal contribution, which involves the decorated Wilson line $\mathcal{U}^{(3)}_{F; ij}$ with $\mathcal{F}^{ij}$ dependence, is obtained as
\begin{align}
& i{\cal M }_{q_1 \bar q_2 \leftarrow \gamma^*_L}^{(3)}
=
 -\frac{Q\, e e_f}{2(q^+)^2}\, \bar u(1)  \gamma^+ \frac{[\gamma^i,\gamma^j]}{4} v(2)
\,
\frac{1}{[\P^2\!+\!\bar Q^2]}  
\int_{\b}\  e^{-i\b\cdot\k}  
\nonumber\\
&\, 
\times\,
\int dz^+ 
\,  t^{a'}
\mathcal{U}_A(+\infty,z^+;\b)_{a'a} \,
g \mathcal{F}_{ij}^a(z^+,\b)\:
+O\left(\frac{Q|\k|}{W^2\, |\P|^3}\right)
\label{ampl-3_L_b2b}
\, .
\end{align}
Finally, the dynamics of the target results in the following form of the subeikonal correction to the amplitude
\begin{align}
& i{\cal M }_{q_1 \bar q_2 \leftarrow \gamma^*_L}^{\textrm{dyn. tar.}}
=
Q\, e e_f\,
 \frac{(z_2\!-\!z_1)}{(q^+)^2}\, \bar u(1)  \gamma^+ v(2)
\,
\frac{[\P^2+m^2]}{[\P^2\!+\!\bar Q^2]^2}  
\int_{\b}\  e^{-i\b\cdot\k}  
\nonumber\\
&\, \times\,
\int dz^+ 
\,  t^{a'}
\mathcal{U}_A(+\infty,z^+;\b)_{a'a} \,
g \mathcal{F}^{+-}_a(z^+,\b)\:
+O\left(\frac{Q|\k|}{W^2\, |\P|^3}\right)
\label{ampl-dyn_tar_L_b2b}
\, .
\end{align}
The three contributions given in Eqs.~\eqref{ampl-1_2_L_b2b_bis}, \eqref{ampl-3_L_b2b} and \eqref{ampl-dyn_tar_L_b2b} obtained in the back-to-back jets regime are the explicit NEik corrections defined in the amplitude~\eqref{NEik_corr_Ampl_L_b2b}.

\section{The cross section of the back-to-back dijet production in DIS}

The eikonal contribution to the cross section for the dijet production via longitudinal photon in the back-to-back regime is found by squaring the strict eikonal amplitude~\eqref{Ampl-Eik_L_b2b} and it gets the form
\begin{align}
&
\frac{d\sigma_{\gamma^{*}_L\rightarrow q_1\bar q_2}}{d {\rm P.S.}}\Bigg|_{\rm Eik}
= 2q^+ \, 2\pi\, \delta(q^+-k^+) 
\,  4\, Q^2\,  e^2 e_f^2\, z_1^3z_2^3
 %
 \int_{\b,\b'}\,   e^{i(\b'-\b)\cdot\k}\,
 \nn \\
 &
 \times
 \Bigg\{
\Bigg[
\frac{4\P^i\P^j}{[\P^2\!+\!\bar Q^2]^4}
-2(z_2\!-\!z_1)\frac{(\P^i\k^j+\k^i\P^j)}{[\P^2\!+\!\bar Q^2]^4}
+16(z_2\!-\!z_1)\frac{(\k\cdot\P)\P^i\P^j}{[\P^2\!+\!\bar Q^2]^5}
\Bigg]
\nn \\
&
\times
 \int  dv^+ d{v'}^+\Big\langle 
 g{\mathcal{F}_{j}^{\;-}}_b ({v'}^+,\b')
 \Big[  \mathcal{U}_A\left(+\infty,{v'}^+;\b'\right)^{\dag}  \mathcal{U}_A\left(+\infty,v^+;\b\right)\Big]_{ba}
  g{\mathcal{F}_{i}^{\;-}}_a (v^+,\b)\Big\rangle
\nonumber\\
&+
\frac{4\P^l}{[\P^2\!+\!\bar Q^2]^4}\bigg[ \delta^{ij}-\frac{4\, \P^i\P^j}{[\P^2\!+\!\bar Q^2]}\bigg]\,
 {\rm tr}_F\big( t^{a'} t^{b'} t^{c'}\big)
 \int  dv^+ dw^+ d{v'}^+
  \Big\langle    \mathcal{U}_A\left(+\infty,{v'}^+;\b'\right)_{c'c}  g{\mathcal{F}_{l}^{\;-}}_c ({v'}^+,\b')
  \nonumber\\
 &
 \times
 \mathcal{U}_A\left(+\infty,{v}^+;\b\right)_{a'a}  g{\mathcal{F}_{i}^{\;-}}_a ({v}^+,\b)
  \mathcal{U}_A\left(+\infty,{w}^+;\b\right)_{b'b}  g{\mathcal{F}_{j}^{\;-}}_b ({w}^+,\b)
  \Big\rangle
 \nonumber\\
 &+
\frac{4\P^l}{[\P^2\!+\!\bar Q^2]^4}\bigg[ \delta^{ij}-\frac{4\, \P^i\P^j}{[\P^2\!+\!\bar Q^2]}\bigg]\,
 {\rm tr}_F\big( t^{a'} t^{b'} t^{c'}\big)
 \int  dv^+ d{w'}^+ d{v'}^+
  \Big\langle   \mathcal{U}_A\left(+\infty,{v}^+;\b\right)_{c'c}  g{\mathcal{F}_{l}^{\;-}}_c ({v}^+,\b)
  \nonumber\\
 &
 \times
 \mathcal{U}_A\left(+\infty,{v'}^+;\b'\right)_{a'a}  g{\mathcal{F}_{i}^{\;-}}_a ({v'}^+,\b')
  \mathcal{U}_A\left(+\infty,{w'}^+;\b'\right)_{b'b}  g{\mathcal{F}_{j}^{\;-}}_b ({w'}^+,\b')
  \Big\rangle
  \Bigg\}+O\left(\frac{Q^2}{\P^8}\right)
  \, .
  \label{str_Eik_X_sec_L}
\end{align}

The first term in the formula (\ref{str_Eik_X_sec_L}) is the leading-power contribution and it is of order $Q^2/(\k^2\, \P^6)$, while all other terms written explicitly are next-to-leading-power corrections, of order $Q^2/(|\k|\, |\P|^7)$. 

Note that if $z^-$ dependence was kept one would get the generalized eikonal cross section of a similar form to the strict eikonal cross section~\eqref{str_Eik_X_sec_L}. These two cross sections would differ by contributions at next-to-next-to-leading-power order in the back-to-back dijet limit, which is beyond the accuracy of the present study. Thus we do not study the generalized eikonal contribution here.

The explicit next-to-eikonal corrections to the cross section, as defined in~\eqref{Cross_Section_long_NEik}, are computed as the interference terms between the eikonal amplitude~\eqref{Ampl-Eik_L_b2b} and next-to-eikonal amplitudes~\eqref{NEik_corr_Ampl_L_b2b}. The final result, where we also include eikonal contribution and then group terms according to types of field strength correlators, can be written in the form 
%
\begin{align}
&
\frac{d\sigma_{\gamma^{*}_L\rightarrow q_1\bar q_2}}{d {\rm P.S.}}\Bigg|_{{\rm Eik }+{\rm NEik}}
= \frac{d\sigma_{\gamma^{*}_L\rightarrow q_1\bar q_2}}{d {\rm P.S.}}\Bigg|_{{\mathcal{F}^{\perp -}} {\mathcal{F}^{\perp -}}}
+\frac{d\sigma_{\gamma^{*}_L\rightarrow q_1\bar q_2}}{d {\rm P.S.}}\Bigg|_{{\mathcal{F}^{\perp -}} {\mathcal{F}^{\perp -}} {\mathcal{F}^{\perp -}}}
+ \frac{d\sigma_{\gamma^{*}_L\rightarrow q_1\bar q_2}}{d {\rm P.S.}}\Bigg|_{{\mathcal{F}^{+ -}} {\mathcal{F}^{\perp -}}}
\, .
\label{X_sec_L_gen_form_a}
\end{align}
%
The first two terms appear as the combination of strict eikonal cross section and the next-to-eikonal interference between the amplitude~\eqref{ampl-1_2_L_b2b_bis} and the strict eikonal amplitude~\eqref{Ampl-Eik_L_b2b}. The third term in Eq.~\eqref{X_sec_L_gen_form_a} is obtained as the interference between the amplitude~\eqref{ampl-dyn_tar_L_b2b} and the strict eikonal amplitude~\eqref{Ampl-Eik_L_b2b}. The three consecutive contributions in the expression~(\ref{X_sec_L_gen_form_a}) are given by
%
%
\begin{align}
&\, \frac{d\sigma_{\gamma^{*}_L\rightarrow q_1\bar q_2}}{d {\rm P.S.}}\Bigg|_{{\mathcal{F}^{\perp -}} {\mathcal{F}^{\perp -}}}
= 2q^+ 2\pi\,  \delta(q^+\!-\!k^+) \, 
\,  4\, Q^2\,  e^2 e_f^2\, z_1^3 z_2^3
\Bigg[
\frac{4\P^i\P^j}{[\P^2\!+\!\bar Q^2]^4}
-2(z_2\!-\!z_1)\frac{(\P^i\k^j+\k^i\P^j)}{[\P^2\!+\!\bar Q^2]^4}
\nonumber\\
 &
 +16(z_2\!-\!z_1)\frac{(\k\!\cdot\!\P)\P^i\P^j}{[\P^2\!+\!\bar Q^2]^5}
\Bigg]
  \int_{\b,\b'}\,   e^{i(\b'-\b)\cdot\k}\, 
   \int  dv^+ d{v'}^+\,
 \bigg[ 1-
 i\frac{[\P^2\!+\!\bar Q^2]}{2q^+z_1 z_2} \, 
 ({v'}^+\!-\!v^+)
 \bigg]
\nonumber\\
&
\times\, 
 \Big\langle 
 g{\mathcal{F}_{j}^{\;-}}_b ({v'}^+,\b')
 \Big[  \mathcal{U}_A\left(+\infty,{v'}^+;\b'\right)^{\dag}  \mathcal{U}_A\left(+\infty,v^+;\b\right)\Big]_{ba}
  g{\mathcal{F}_{i}^{\;-}}_a (v^+,\b)\Big\rangle
 +O\left(\frac{Q^2}{\P^8}\right)
 +O\left(\frac{Q^2}{\k^2 \P^2 W^4}\right)
\label{X_sec_L_Fperpmin_Fperpmin}
\end{align}
%
and
%
%
\begin{align}
&
\frac{d\sigma_{\gamma^{*}_L\rightarrow q_1\bar q_2}}{d {\rm P.S.}}\Bigg|_{{\mathcal{F}^{\perp -}} {\mathcal{F}^{\perp -}} {\mathcal{F}^{\perp -}}}
= 2q^+ 2\pi\,  \delta(q^+-k^+) \, 16\, Q^2\,  e^2 e_f^2\, z_1^3 z_2^3\, \frac{\P^l}{[\P^2\!+\!\bar Q^2]^4}\, 
\nn \\
&    
2\, {\textrm{Re}}\;    
\,  
 \int_{\b,\b'}\,   e^{i(\b'-\b)\cdot\k}\, 
  \int  dv^+ dw^+ d{v'}^+
 {\rm tr}_F\big( t^{a'} t^{b'} t^{c'}\big)\;
 \Big\langle   \mathcal{U}_A\left(+\infty,{v'}^+;\b'\right)_{c'c}  g{\mathcal{F}_{l}^{\;-}}_c ({v'}^+,\b')
 \nn \\
 &
 \times
 \mathcal{U}_A\left(+\infty,{v}^+;\b\right)_{a'a}  g{\mathcal{F}_{i}^{\;-}}_a ({v}^+,\b)
  \mathcal{U}_A\left(+\infty,{w}^+;\b\right)_{b'b}  g{\mathcal{F}_{j}^{\;-}}_b ({w}^+,\b)
  \Big\rangle
   \nonumber\\
 &
 \times
 \Bigg\{
\delta^{ij}\bigg[1-  i\frac{[\P^2\!+\!\bar Q^2]}{2q^+z_1 z_2} \, 
 \left({v'}^+\!-\!\min(v^+,w^+)\right)\bigg]
-\frac{4\, \P^i\P^j}{[\P^2\!+\!\bar Q^2]}
\bigg[1-  i\frac{[\P^2\!+\!\bar Q^2]}{2q^+z_1 z_2} \, 
 \left({v'}^+\!-\!z_2 v^+\!-\!z_1 w^+)\right)\bigg]
   \Bigg\}
  \nonumber\\
 &  
   +O\left(\frac{Q^2}{\P^8}\right)
 +O\left(\frac{Q^2}{\k^2 \P^2 W^4}\right)
   \, ,
\label{X_sec_L_Fperpmin_Fperpmin_Fperpmin}   
\end{align}
and
\begin{align}
&\, \frac{d\sigma_{\gamma^{*}_L\rightarrow q_1\bar q_2}}{d {\rm P.S.}}\Bigg|_{{\mathcal{F}^{+ -}} {\mathcal{F}^{\perp -}}}
=  2q^+ \, 2\pi\, \delta(q^+-k^+) 
\,  8\, Q^2\,  e^2 e_f^2\, \frac{z_1^2z_2^2 (z_2\!-\!z_1)}{q^+} \; \frac{\P^j[\P^2+m^2]}{[\P^2\!+\!\bar Q^2]^4}\,
 2\, {\textrm{Re}}\,  \int_{\b,\b'}\,   e^{i(\b'-\b)\cdot\k}\,
\nn \\
&
\times
 \int  dv^+ d{v'}^+
 \Big\langle 
 g{\mathcal{F}_{j}^{\;-}}_b ({v'}^+,\b')
 \Big[  \mathcal{U}_A\left(+\infty,{v'}^+;\b'\right)^{\dag}  \mathcal{U}_A\left(+\infty,v^+;\b\right)\Big]_{ba}
g\mathcal{F}^{\, +-}_a (v^+,\b)\Big\rangle
\nn \\
 &
+O\left(\frac{Q^2}{\P^6 W^2}\right)
+O\left(\frac{Q^2}{|\k|\, |\P|^3 W^4}\right)
  \, .
\label{X_sec_L_Fplusmin_Fperpmin}   
\end{align}
%
Let us point out that the term~\eqref{ampl-3_L_b2b} in the amplitude does not contribute to the cross section at next-to-eikonal accuracy because its Dirac structure vanishes
%
%
\begin{align}
\sum_{h_1,h_2}\, \bar v(2) \gamma^+ u(1)\,\bar u(1) \gamma^+ [\gamma^i,\gamma^j] v(2)=0
\, .
\end{align}
%

An essential observation in our results is that the next-to-leading-power and the next-to-eikonal power corrections factorize from each other in Eq.~\eqref{X_sec_L_Fperpmin_Fperpmin}. In that expression, the next-to-leading-power corrections are purely kinematical corrections to the hard factor, while the next-to-eikonal correction (involving the light-cone time interval ${v'}^+\!-\!v^+$ between the two field strength insertions) is a correction to the operator.

To connect the cross section given by the expression~(\ref{X_sec_L_gen_form_a}) with the TMD results we introduce the following notation for the gluon field strength correlator in an unpolarized target
%
\begin{align}
\Phi^{\mu\nu;\rho\sigma}({\rm x},\k) \equiv&\,
\frac{1}{{\rm x} P_{{tar}}^-}\, \frac{1}{(2\pi)^3}
 \int d^2\z\,   e^{-i\k \cdot\z}\, 
   \int  d z^+\, e^{i {\rm x} P_{{tar}}^- z^+}\, 
\nonumber\\
 &\, \times\, 
  \Big\langle P_{{tar}}\Big|
 \mathcal{F}^{\mu\nu}_b (0)
 \Big[  \mathcal{U}_A\left(+\infty,0;0\right)^\dagger  \mathcal{U}_A\left(+\infty,z^+;\z\right)\Big]_{ba}
  \mathcal{F}^{\rho\sigma}_a (z)\Big| P_{{tar}}\Big\rangle \bigg|_{z^-=0}
\label{def_Phi_correlator_Fmunu_Frhosigma}
\, ,
\end{align}
with a gauge link forming a future pointing staple. The angle brackets in the correlator (\ref{def_Phi_correlator_Fmunu_Frhosigma}) denote the expectation value of an operator in the quantum state of the target
(see for example Refs.~\cite{Dominguez:2011wm,Altinoluk:2019wyu}, and should be understood as
\ba
\langle{\cal O} \rangle &=& \lim_{P_{{tar}}'\rightarrow P_{{tar}}}
\frac{\langle P_{{tar}}'|\hat{\cal O}| P_{{tar}}\rangle}{\langle P_{{tar}}'| P_{{tar}}\rangle}
\, ,
\label{def_average}
\ea
where the target states are normalized as
\ba
\langle P_{{tar}}'| P_{{tar}}\rangle &=& 2P_{{tar}}^-\, (2\pi)^3 \delta(P_{{tar}}^{'-}\!-\!P_{{tar}}^-)\, \delta^{(2)}(\P_{{tar}}'\!-\!\P_{{tar}})
\, .
\label{norm_states}
\ea

Using the notation (\ref{def_Phi_correlator_Fmunu_Frhosigma}), the cross section~\eqref{X_sec_L_gen_form_a} can be rewritten in the form 
%
\begin{align}
\frac{d\sigma_{\gamma^{*}_{L}\rightarrow q_1\bar q_2}}{d {\rm P.S.}}\Bigg|_{{\rm Eik }+{\rm NEik}} \!\!\!
= &\,
2q^+ \, 2\pi\, \delta(q^+\!-\!k^+) 
\,  z_1 z_2\,  e^2 e_f^2\, g^2 \; \frac{(2\pi)^3}{2}\,
\nn \\
& \,
\times
\Bigg\{
{\cal C}^{ij}_{L}(z_1,\P,\k)
 \bigg[ 1 +\frac{[\P^2\!+\!\bar Q^2]}{z_1 z_2(2q^+P_{{tar}}^-)} \, \partial_{{\rm x}}\bigg]  
 \bigg[ {\rm x}\, \Phi^{j -;i-}({\rm x},\k) \bigg] 
 \nn\\
 &\,
 -\frac{1}{2q^+}\, {\cal C}^{j}_{L}(z_1,\P)\bigg[  {\rm x}\, \Phi^{j-;+-}({\rm x},\k) + {\rm x}\, \Phi^{+-;j-}({\rm x},\k)\bigg]
  \nn\\
 &\,
 -\frac{1}{2q^+}\, {\cal C}^{ijl}_{L}(z_1,\P)\bigg[ {\rm x}\, \Phi^{l-;ij}({\rm x},\k)  + {\rm x}\, \Phi^{ij;l-}({\rm x},\k)\bigg]
\Bigg\}\Bigg|_{{\rm x}=0}
+3\, \mathcal{F}\textrm{ terms}
\, ,
\label{X_sec_gen_form_Phi}
\end{align}
where the hard factors ${\cal C}^{ij}_{L}$, ${\cal C}^{j}_{L}$ and ${\cal C}^{ijl}_{L}$ which can be read off from the expressions (\ref{X_sec_L_Fperpmin_Fperpmin}-\ref{X_sec_L_Fplusmin_Fperpmin}).

The second line of Eq.~\eqref{X_sec_gen_form_Phi} contains the leading operator structure $\Phi^{j -;i-}$, and the coefficient ${\cal C}^{ij}_{L}$ contains next-to-leading-power (kinematical twist) corrections and the expression in the $[1+\dots]$ bracket contains next-to-eikonal corrections. As mentioned previously, the two types of the corrections factorize from each other. The two last lines feature new operator structures that are purley next-to-eikonal corrections. The $3\, \mathcal{F}$ term contains eikonal next-to-leading power (genuine twist) corrections and other contributions that are both next-to-leading-power and subeikonal. 

The field strength correlators of the type \eqref{def_Phi_correlator_Fmunu_Frhosigma} are usually parametrized in terms of gluon TMD distributions that are real scalar functions as
%
\begin{align}
 \Phi^{j -;i-}({\rm x},\k) =&\,
\frac{\delta^{ij}}{2}\, f_1^g({\rm x},\k) +\left[\k^i\k^j-\frac{\k^2}{2}\, \delta^{ij}\right]\frac{1}{2 M^2}\, h_1^{\perp g}({\rm x},\k)
\nn\\
\Phi^{j-;+-}({\rm x},\k) =&\,
\frac{\k^j}{P_{{tar}}^-}\, \left[f^{\perp g}({\rm x},\k)-i  \bar{f}^{\perp g}({\rm x},\k)\right] 
\end{align}
\begin{align}
\Phi^{+-;j-}({\rm x},\k) =&\,
\frac{\k^j}{P_{{tar}}^-}\, \left[f^{\perp g}({\rm x},\k)+i  \bar{f}^{\perp g}({\rm x},\k)\right] 
\nn\\
\Phi^{l-;ij}({\rm x},\k) =&\,
\epsilon^{ij}\, \epsilon^{ln}\, 
\frac{\k^n}{P_{{tar}}^-}\, \left[ \bar{g}^{\perp g}({\rm x},\k)+i  g^{\perp g}({\rm x},\k)\right] 
\nn\\
\Phi^{ij;l-}({\rm x},\k) =&\,
\epsilon^{ij}\, \epsilon^{ln}\, 
\frac{\k^n}{P_{{tar}}^-}\, \left[ \bar{g}^{\perp g}({\rm x},\k)-i  g^{\perp g}({\rm x},\k)\right] 
\label{param_Phi}
\, ,
\end{align}
where $M$ is the target mass.
Applying the relations \eqref{param_Phi} to  Eq.~\eqref{X_sec_gen_form_Phi} one obtains
%
\begin{align}
\frac{d\sigma_{\gamma^{*}_{L}\rightarrow q_1\bar q_2}}{dz_1\, d^2\P\, d^2\k}\Bigg|_{{\rm Eik }+{\rm NEik}}
= &\,
  \alpha_{\rm em} e_f^2\, \alpha_{s} \; 
\Bigg\{
{\cal C}^{ f_1^g}_{L}(z_1,\P,\k)\:
 {\rm x}\, f_1^g({\rm x},\k) 
 + {\cal C}^{ h_1^{\perp g}}_{L}(z_1,\P,\k)\:
{\rm x}\, h_1^{\perp g}({\rm x},\k)  
 \nn\\
 &
+\frac{\k\!\cdot\!\P}{W^2}\, {\cal C}^{ f^{\perp g}}_{L}(z_1,\P)\:  {\rm x}\, f^{\perp g}({\rm x},\k)
+\frac{\k\!\cdot\!\P}{W^2}\,  {\cal C}^{  \bar{g}^{\perp g}}_{L}(z_1,\P)\:
{\rm x}\,  \bar{g}^{\perp g}({\rm x},\k)
\Bigg\}\Bigg|_{{\rm x}=\frac{[\P^2\!+\!\bar Q^2]}{z_1 z_2\, W^2} \,} \nn \\
&
+3\, \mathcal{F}\textrm{ terms}
\label{X_sec_gen_form_scal_TMDs_x_non_zero}
\end{align}
where we used the freedom to exponentiate the relevant terms into the x dependent phases, within the same accuracy in our beyond-eikonal expansion. The coefficients are given by
%

\begin{align}
 {\cal C}^{ f_1^g}_{L}(z_1,\P,\k) 
\equiv &\,
\frac{8Q^2\, z_1^2 z_2^2}{[\P^2\!+\!\bar Q^2]^4} 
\left\{\P^2 +(z_2\!-\!z_1)(\k\!\cdot\!\P) \left[-1
+\frac{4\P^2}{[\P^2\!+\!\bar Q^2]}\right]
\right\}
\nn\\
 {\cal C}^{ h_1^{\perp g}}_{L}(z_1,\P,\k) 
\equiv &\,
 \frac{4Q^2\, z_1^2 z_2^2}{[\P^2\!+\!\bar Q^2]^4}\, \frac{\k^2}{M^2} 
\Bigg\{\left(\frac{2(\k\!\cdot\!\P)^2}{\k^2\P^2}-1\right)\P^2 
\nn \\
&
+(z_2\!-\!z_1)(\k\!\cdot\!\P) \left[-1
+\frac{4\P^2}{[\P^2\!+\!\bar Q^2]}\left(\frac{2(\k\!\cdot\!\P)^2}{\k^2\P^2}-1\right)\right]
\Bigg\}
\nn\\
{\cal C}^{ f^{\perp g}}_{L}(z_1,\P) 
\equiv &\,
 -32Q^2\, z_1 z_2\,
 \frac{(z_2\!-\!z_1)[\P^2\!+\!m^2]}{ [\P^2\!+\!\bar Q^2]^4}
 \nn\\
{\cal C}^{  \bar{g}^{\perp g}}_{L}(z_1,\P) 
\equiv &\,
0
\label{scal_coeff_contract_L}
\end{align}
Summarizing, the dependence on ${\rm x}$, characteristic of the TMDs, in the formula (\ref{X_sec_gen_form_scal_TMDs_x_non_zero}) was recovered in the high-energy formalism by resumming a subset of power corrections obtained beyond the eikonal approximation. 

\section*{Acknowledgements}

TA is supported in part by the National Science Centre (Poland) under the research Grant No. 2023/50/E/ST2/00133 (SONATA BIS 13). GB is supported in part by the National Science Centre (Poland) under the research Grant No. 2020/38/E/ST2/00122 (SONATA BIS 10). AC is supported in part by the National Science Centre (Poland) under the research Grant No. 2021/43/D/ST2/01154 (SONATA 17).

\end{document}